\documentclass[preprint,12pt,numafflabel]{elsarticle}

\graphicspath{{Figures/}} 
\usepackage{amsmath}      
\usepackage{xcolor}       
\usepackage{multicol}
\usepackage{soul}
\usepackage{url}
\usepackage{comment}
\usepackage[colorlinks=true]{hyperref} 

\usepackage{graphicx}
\usepackage{amssymb}

\usepackage{siunitx}	

\biboptions{sort&compress}

\journal{Nature Communications}

\newcounter{daggerfootnote}
\newcommand*{\daggerfootnote}[1]{%
    \setcounter{daggerfootnote}{\value{footnote}}%
    \renewcommand*{\thefootnote}{\fnsymbol{footnote}}%
    \footnote[2]{#1}%
    \setcounter{footnote}{\value{daggerfootnote}}%
    \renewcommand*{\thefootnote}{\arabic{footnote}}%
    }

\makeatletter
\def\ps@pprintTitle{%
  \let\@oddhead\@empty
  \let\@evenhead\@empty
  \let\@oddfoot\@empty
  \let\@evenfoot\@oddfoot
}
\makeatother

\begin{document}

\begin{frontmatter}

\title{Search for Dark Matter Axions with CAST-CAPP}


\author[vancouver]{    C.~M.~Adair}
\author[zaragoza]{    K.~Altenm\"{u}ller}
\author[patras]{    V.~Anastassopoulos} 
\author[cern]{ S.~Arguedas Cuendis}
\author[freiburg]{    J.~Baier} 
\author[cern]{    K.~Barth} 
\author[inr]{    A.~Belov  }
\author[rijekaeng]{  D.~Bozicevic }
\author[mpe]{    H.~Br\"auninger  } 
\author[trieste,triesteu]{    G.~Cantatore  }  
\author[esi,cern]{F.~Caspers}
\author[zaragoza]{    J.~F.~Castel  }  
\author[istinye]{    S.~A.~\c{C}etin  }
\author[capp]{     W.~Chung  }
\author[kaist]{     H.~Choi  }
\author[capp]{     J.~Choi  }
\author[zaragoza]{    T.~Dafni  }  
\author[cern]{    M.~Davenport  }  
\author[inr]{    A.~Dermenev  }  
\author[bonn]{	K.~Desch}  
\author[cern]{    B.~D\"obrich  } 
\author[freiburg]{    H.~Fischer  }  
\author[cern]{    W.~Funk  }  
\author[zaragoza]{    J.~Galan}
\author[patras,hamburg]{    A.~Gardikiotis  } 
\author[inr]{    S.~Gninenko  }  
\author[cern,jena]{    J.~Golm  } 
\author[vancouver]{    M.~D.~Hasinoff  }  
\author[xian]{    D.~H.~H.~Hoffmann  } 
\author[zaragoza]{   D.~D\'{\i}ez Ib\'a\~{n}ez}  
\author[zaragoza]{   I.~G.~Irastorza  }  
\author[zagreb]{   K.~Jakov\v ci\' c  }  
\author[bonn]{	J.~Kaminski}  
\author[rijeka,cerijeka,trieste]{   M.~Karuza } 
\author[bonn,x3]{ 	C. Krieger} 
\author[kaist,capp]{     \c{C}.~Kutlu  }
\author[zagreb]{  B.~Laki\'{c} \daggerfootnote{deceased}}  
\author[cern]{    J.~M.~Laurent  } 
\author[kaist]{    J.~Lee  }
\author[capp]{   S.~Lee  }
\author[zaragoza]{    G.~Luz\'on  }
\author[cern]{   C.~Malbrunot}
\author[zaragoza]{    C.~Margalejo  } 
\author[patras]{    M.~Maroudas* }  
\author[capp]{    L.~Miceli  }  
\author[zaragoza]{    H.~Mirallas  } 
\author[zaragoza]{    L.~Obis}
\author[istinye,x4]{A.~\"{O}zbey}
\author[istinye,x1]{    K.~\"{O}zbozduman*  }  
\author[llnl,x5]{    M.~J.~Pivovaroff  }
\author[eli]{    M.~Rosu  }
\author[llnl]{    J.~Ruz  } 
\author[mainz]{    E.~Ruiz-Ch\'oliz  }  
\author[bonn]{    S.~Schmidt  } 
\author[freiburg]{M.~Schumann }
\author[capp,kaist]{    Y.~K.~Semertzidis  }
\author[mpis]{    S.~K.~Solanki  }
\author[cern]{    L.~Stewart  }  
\author[patras]{    I.~Tsagris} 
\author[cern]{    T.~Vafeiadis  } 
\author[llnl]{    J.~K.~Vogel  } 
\author[rijeka,x6]{    M.~Vretenar  }  
\author[capp]{    S.~Youn  }
\author[patras,cern]{    K.~Zioutas  } 
%
\address[vancouver]{Department of Physics and Astronomy, University of British Columbia, V6T 1Z1 Vancouver, Canada}
\address[zaragoza]{Centro de Astropart\'{\i}culas y F\'{\i}sica de Altas Energ\'{\i}as (CAPA), Universidad de Zaragoza, 50009 Zaragoza, Spain}
\address[patras]{Physics Department, University of Patras, 26504 Patras, Greece}
\address[cern]{European Organization for Nuclear Research (CERN), CH-1211 Gen\`eve, Switzerland}
\address[freiburg]{Physikalisches Institut, Albert-Ludwigs-Universit\"{a}t Freiburg, 79104 Freiburg, Germany}
\address[inr]{Institute for Nuclear Research (INR), Russian Academy of Sciences, 117312 Moscow, Russia}
\address[rijekaeng]{University of Rijeka, Faculty of Engineering, 51000 Rijeka, Croatia}
\address[mpe]{Max-Planck-Institut f\"{u}r Extraterrestrische Physik, D-85741 Garching, Germany}
\address[trieste]{Istituto Nazionale di Fisica Nucleare (INFN), Sezione di Trieste, 34127 Trieste, Italy} 
\address[triesteu]{Universit\`a di Trieste, 34127 Trieste, Italy}
\address[esi]{European Scientific Institute (ESI), 74160 Archamps, France}
\address[istinye]{Istinye University, Institute of Sciences, 34396 Sariyer, Istanbul, T\"{u}rkiye}
\address[capp]{Center for Axion and Precision Physics Research, Institute for Basic Science (IBS), Daejeon 34141, Republic of Korea}
\address[kaist]{Department of Physics, Korea Advanced Institute of Science and Technology (KAIST), Daejeon 34141, Republic of Korea}
\address[bonn]{Physikalisches Institut, University of Bonn, 53115 Bonn, Germany}
\address[hamburg]{Universit\"{a}t Hamburg, 22762 Hamburg, Germany}
\address[jena]{Institute for Optics and Quantum Electronics, Friedrich Schiller University Jena, 07743 Jena, Germany}
\address[xian]{Xi'An Jiaotong University, School of Science, Xi'An, 710049, China}
\address[zagreb]{Rudjer Bo\v{s}kovi\'{c} Institute, 10000 Zagreb, Croatia}
\address[rijeka]{University of Rijeka, Faculty of Physics, 51000 Rijeka, Croatia}
\address[cerijeka]{University of Rijeka, Photonics and Quantum Optics Unit, Center of Excellence for Advanced Materials and Sensing Devices, and Centre for Micro and Nano Sciences and Technologies, 51000 Rijeka, Croatia}
\address[llnl]{Lawrence Livermore National Laboratory, Livermore, CA 94550, USA}
\address[eli]{Extreme Light Infrastructure - Nuclear Physics (ELI-NP), 077125 Magurele, Romania}
\address[mainz]{Institut f\"{u}r Physik, Johannes Gutenberg Universit\"{a}t Mainz, 55128 Mainz, Germany}
\address[mpis]{Max-Planck-Institut f\"{u}r Sonnensystemforschung, 37077 G\"{o}ttingen, Germany}

%
\address[x1]{Present address: Bogazici University, Physics Department, 34342 Bebek, Istanbul, T\"{u}rkiye}
\address[x3]{Present address: Institute of Experimental Physics, University of Hamburg, 22761 Hamburg, Germany}
\address[x4]{Present address: Istanbul University - Cerrahpasa, Department of Mechanical Engineering, 34320 Istanbul, T\"{u}rkiye}
\address[x5]{Present address: SLAC National Accelerator Laboratory, Menlo Park, CA 94025, U.S.A.}
\address[x6]{Present address: Adaptive Quantum Optics (AQO), MESA+Institute for Nanotechnology, University of Twente, PO Box 217, 7500 AE Enschede, The Netherlands}

\cortext[mycorrespondingauthor]{Corresponding authors: \\ marios.maroudas@cern.ch, kaan.ozbozduman@cern.ch}
%

\begin{abstract}

The CAST-CAPP axion haloscope, operating at CERN inside the CAST dipole magnet, has searched for axions in the \SIrange{19.74}{22.47}{\micro\eV} mass range. The detection concept follows the Sikivie haloscope principle, where Dark Matter axions convert into photons within a resonator immersed in a magnetic field. The CAST-CAPP resonator is an array of four individual rectangular cavities inserted in a strong dipole magnet, phase-matched to maximize the detection sensitivity. Here we report on the data acquired for \SI{4124}{\hour} from 2019 to 2021. Each cavity is equipped with a fast frequency tuning mechanism of \SI{10}{\MHz}\,/\,min between \SI{4.774}{\GHz} and \SI{5.434}{\GHz}. In the present work, we exclude axion-photon couplings for virialized galactic axions down to $ g_{a\gamma\gamma} = 8 \times 10^{-14}\,\si{\GeV\tothe{-1}} $ at the $90\%$ confidence level. The here implemented phase-matching technique also allows for future large-scale upgrades.
\end{abstract}


\end{frontmatter}


\section{Introduction}
\label{S:1}

The quest to understand and identify the nature of Dark Matter (DM) is gaining considerable momentum, and within this framework experimental searches for DM axions have come to prominence.
Axions are nearly massless pseudoscalar bosons that appear in many extensions of the standard model \cite{PecceiQuinn:1977prd, Zhitnitsky:1980sjnp, Sikivie:1983prl, Kim:1979prl, Shifman:1980npb, Dine:1981plb}. Of particular interest is the case of the quantum chromodynamics (QCD) axion, which arises from the Peccei–Quinn mechanism proposed as a solution to the ``strong CP problem" in QCD \cite{PecceiQuinn:1977prd}. Because of their extremely feeble interactions DM axions are often dubbed ``invisible" and so far most search strategies rely on their coupling to an external magnetic field and their conversion into photons following the Lagrangian

\begin{equation}
\mathcal{L}_{a\gamma\gamma} = g_{a\gamma\gamma}\, a\, \mathbf{E} \cdot \mathbf{B},
    \label{eq:lagrangian}
\end{equation}

\noindent
where $g_{a\gamma\gamma} = \alpha g_{\gamma} / 2 \pi f_a$ is the axion-photon coupling coefficient, $a$ the axion field, $\mathbf{E} \cdot \mathbf{B}$ the scalar product of the electric and magnetic field strength, $\alpha$ the fine structure constant, $f_a$ the axion decay constant and $g_{\gamma}$ a model dependent dimensionless coupling constant \cite{Sikivie:1983prl}. Most widely quoted are the KSVZ (Kim-Shifman-Vainshtein-Zakharov) \cite{Kim:1979prl,Shifman:1980npb} and DFSZ (Dine-Fischler-Srednicki-Zhitnitsky) \cite{Zhitnitsky:1980sjnp,Dine:1981plb} models for which $g_{\gamma}$ varies between $-0.97$ and $0.36$. 

The main search method, used for over 30 years, is based on this coupling and employs the concept of the Sikivie haloscope \cite{Sikivie:1983prl}, searching for DM axions converting to photons inside a high-quality resonant cavity immersed in a strong magnetic field. The quality factor $Q_{\text{L}}$ of the cavity enhances the conversion probability, provided that the axion mass matches the resonant frequency of the cavity. Since the axion mass is unknown, haloscope searches must be conducted by tuning the cavity in successive steps to scan the expected axion mass range.
The scanning speed is a function of several parameters, but is mainly determined by choosing a suitably long integration time such that at a given frequency the axion haloscope achieves sufficient sensitivity to probe axion model predictions. Haloscope searches require an a priori axion distribution model \cite{gianotti_2020}, and the common approach is to assume an isotropic density of DM axions \cite{read_2014}. Under this assumption, relatively long integration times are required for each step, while an assumption of concentrated streams or flows of DM axions leads to reduced integration times.

The CERN Axion Solar Telescope (CAST) has been operating for two decades as an experimental platform for axions and other exotic particles. Originally developed as an axion helioscope capable of detecting axions produced in the Sun, CAST has undergone several steps of upgrades to improve its sensitivity for axions, axion-like particles (ALPs) and chameleons \cite{cast-alps-2010, CAST-2017, cast-chameleons-2015, cast-gridpix-2019, cast-kwisp-2019, cast-rades-2021}. With this new science program, CAST becomes unique amongst all axion experiments employing the helioscope and haloscope techniques following a suggestion from 2012 \cite{baker-2012}. Several features were implemented for the first time in an axion haloscope ``\`a la Sikivie": a dipole magnet with a high degree of field homogeneity, four phase-matched resonant cavities, and a fast frequency scanning technique. The fast scanning technique could become decisive in the discovery of the DM axion, taking into account possible DM axion transients such as streams \cite{vogelsberger-2011} and Mini Clusters \cite{Tkachev-1991, Kolb-1993}. Gravitational lensing effects of aligned DM streams \cite{hoffmann-2003, patla-2013} by solar system bodies can temporally increase the signal strength significantly \cite{zioutas-2017, kryemadhi-gravitational, fischer-search}. In this work we report on the search for galactic DM axions while the here introduced detection technique adapted for transient events will be the subject of a forthcoming paper.

\section{Results}
\label{S:2}

\subsection{Materials and methods}
\label{S:2_1}

DM axions with velocity $v \sim 10^{-3}c$, where $c$ is the speed of light, have negligible kinetic energy $\mathcal{O}(10^{-6})$ compared to their rest mass. Thus, the frequency of the axion-converted photons is $\nu_a \approx m_ac^2/h$, corresponding to the microwave domain for $m_a \sim \si{\micro\eV}$. Microwave photons converted from axions resonate inside a cavity when the photon frequency matches that of the cavity resonance mode. The resonance mode frequencies $\nu$ for a rectangular cavity are given by:

\begin{equation}
    \nu_{lmn} = \frac{c}{2} \sqrt{ \left( \frac{l}{L_x} \right)^2 + \left( \frac{m}{L_y} \right)^2 + \left( \frac{n}{L_z} \right)^2 }
    \label{eq:cavity-frequency}
\end{equation}

\noindent
where $L_x, L_y, L_z$ are cavity side lengths and $l,m,n$ are the corresponding mode indices which are defined by the separation condition. A tuning mechanism inside the cavity is used to alter the boundary conditions and change the electromagnetic field solutions to scan a range of photon frequencies, i.e., axion masses. Not all resonance modes are axion sensitive because axion-photon conversion requires the magnetic field to overlap the electric field of the resonance mode. This condition favors transverse electric (TE) modes where the radio frequency (RF) electric field is parallel to the static magnetic field which fits the CAST-CAPP design. The selection of the TE mode is through the maximization of the cavity form factor:
\begin{equation}
    C_{lmn} = \frac{\left| \int dV \ \mathbf{E}_{lmn}(x,y,z) \cdot \mathbf{B} \right|^2}{B^2 V \int dV \ \varepsilon(x,y,z) \left| \textbf{E}_{lmn}(x,y,z) \right|^2 }
    \label{eq:geometry_factor}
\end{equation}
which is a dimensionless number between 0 and 1 as a measure of the overlap of resonant electric and static magnetic fields. The optimum mode for axion searches in CAST-CAPP is TE$_{101}$ as it maximizes the form factor around 5 GHz at $C_{101} = 0.53 \pm 0.05$, computed by numerical field simulations, where the variation on the form factor is due to frequency-dependence. Here, $V$ and $\mathbf{B}$ are the cavity volume and the static magnetic field value, $E_{lmn}$ is the electric field associated with the resonance mode and $\varepsilon(x,y,z)$ is the permittivity distribution inside the cavity.

Axion-photon conversion inside a resonant cavity can result in a voltage signal at a cavity coupler. The power of the converted axion signal at resonance for a single CAST-CAPP cavity is given by:

\begin{align}
    P_{\text{axion}} =& 2.1\times10^{-24}\hspace{0.1cm} \text{W} \times \left( \frac{\beta}{1 + \beta}\right) \left( \frac{V}{224 \hspace{0.1cm} \text{cm}^3} \right) \left( \frac{B}{8.8 \hspace{0.1cm} \text{T}}\right)^2 \left( \frac{C_{lmn}}{0.53}\right)
    \nonumber \\
    &\times \left( \frac{g_{\gamma}}{0.97} \right)^2 \left( \frac{\rho}{0.45 \hspace{0.1cm} \text{GeV}/\text{cm}^3} \right)
    \left( \frac{Q_{\text{L}}}{2 \times 10^4} \right) \left( \frac{\nu_a}{5 \hspace{0.1cm} \text{GHz}} \right) 
    \nonumber \\
    &\times \left( 1 + (2 Q_{\text{L}} (\nu-\nu_0)/\nu_0)^2 \right)^{-1}
    \label{eq:axion-power}
\end{align}

\noindent
The denominator values in most of the terms in Eq.~\eqref{eq:axion-power} reflect CAST-CAPP experimental parameters: $\beta$ is the coupling of the signal output coupler, $g_{\gamma}$ is the model-dependent dimensionless coupling constant. The local DM density $\rho$ is generally accepted to be about 0.45 GeV/cm$^3$ which corresponds to $\sim 10^{13}$ axion/cm$^3$ in our axion mass range \cite{read_2014, ADMX-2001} if the axion is the only DM constituent. The loaded quality factor of the cavity $Q_{\text{L}}$ is the ratio of the resonance mode frequency to the \SI{3}{\dB} points of the resonance mode as measured by a Vector Network Analyzer (VNA), which in our case is $Q_{\text{L}} \approx 2 \times 10^4$(at around \SI{5.3}{\GHz}), at cryogenic conditions with magnetic field. The measured $Q_{\text{L}}$ variation between the four cavities for the same frequency and in cryogenic conditions is below $10\%$, whereas the difference is $5\%$ with or without static magnetic dipole field. The uncertainty in each specific measurement of $Q_{\text{L}}$ is $< \pm 1.5\%$.

The resonance mode has a Lorentzian profile that affects the axion-photon conversion power via the last term of Eq.~\eqref{eq:axion-power} where $\nu_0$ is the center frequency of the resonance. $f_{\text{Lab}}(\nu)$ is the probability distribution function of the axion lineshape in lab frame \cite{HAYSTAC-2017}:
\begin{equation}
    f_{\text{Lab}}(\nu) =
    \frac{c^2\sqrt{6/\pi}}{r\nu_a \langle v^2 \rangle}
    \exp \left( -\frac{3r^2}{2} -\frac{3c^2(\nu-\nu_a)}{\nu_a \langle v^2 \rangle} \right)
    \text{sinh} \left( 3r \sqrt{\frac{2c^2(\nu-\nu_a)}{\nu_a \langle v^2 \rangle}} \right)
    \label{eq:axion-pdf}
\end{equation}
where r $\approx$ 0.85 is the ratio of the velocity of the Earth with respect to the galactic rest frame to RMS velocity of galactic halo.
The DM halo of the Milky Way is virialized with RMS velocity $\sqrt{\langle v^2 \rangle} \approx 270$ km/s being Maxwell-Boltzmann distributed (Eq.~\eqref{eq:axion-pdf}) with the quality factor $Q_{\text{axion}} \sim 10^6$. Hence the axion signal linewidth is $\Delta \nu_a \approx5$ kHz for axion mass $\sim20 \; \mu$eV. These variables are modified for the laboratory frame considering the motion of the Sun and the Earth through the galactic halo which yield $\sqrt{\langle v^2 \rangle} \approx 350$ km/s and linewidth $\Delta \nu_a\approx7$ kHz \cite{Turner-1990}.

The signal-to-noise ratio (SNR) of an axion haloscope is a figure of merit reflecting the sensitivity of the axion antenna and also determines the tuning speed given a target sensitivity for the axion-photon coupling constant $g_{a\gamma\gamma}$:

\begin{equation}
    \text{SNR} = \frac{P_{\text{axion}}}{\sigma_{\text{noise}}} = \frac{P_{\text{axion}}}{k T_{\text{sys}}} \sqrt{\frac{t}{\Delta \nu_a}} 
    \label{eq:snr}
\end{equation}

\noindent
where $k$ is the Boltzmann constant, and $\sigma_{\text{noise}}$ is the standard deviation of the mean noise power $P_{\text{noise}} = k T_{\text{sys}} \Delta \nu$, which is reduced by a longer data integration time $t$. The system temperature $T_{\text{sys}}$ is the sum of the physical cavity temperature and the electronic noise temperature of the receiver. 

For a DM axion of $m_a \sim \SI{20}{\micro\eV}$, the minimum coherence length is $\sim \SI{60}{\meter}$ which is much larger than the size of the cavity setup of $\sim \SI{2}{\meter}$ length. 

CAST-CAPP consists of four tuneable $23 \times 25 \times 390$\si{\mm} rectangular cavities. The volume of each cavity is $V = \SI{224}{\centi\meter\cubed}$. Each cavity consists of two pieces of stainless steel, coated with \SI{30}{\micro\meter} of copper. They are installed in series inside one of the two bores, \SI{43}{\milli\meter} $\varnothing$, of CAST's superconducting dipole magnet at CERN with the split plane parallel to the magnetic field along the cavity's small face. The CAST dipole magnet is an LHC prototype and has a \SI{9.25}{\meter} long uniform magnetic field of \SI{8.8}{\tesla}.

\begin{figure}[!htb]
\centering
\includegraphics[width=0.9\linewidth]{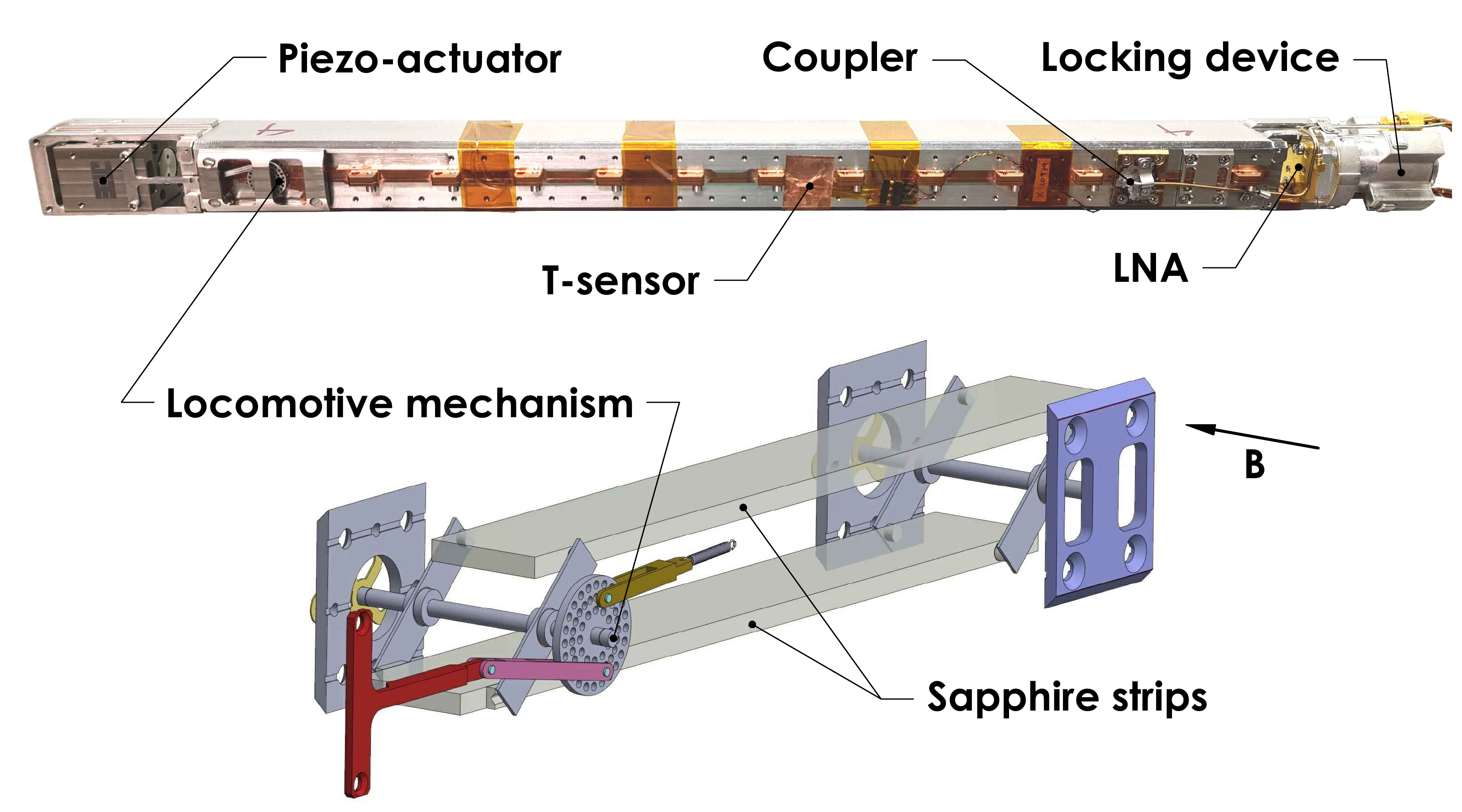}
\caption{A photograph of the elements of a single cavity assembly (top) and a technical drawing of CAST-CAPP tuning mechanism with the two sapphire strips (bottom). The static B-field is shown by the arrow and is parallel to the two axes of the tuning mechanism.}
\label{fig:cavity_assembly}
\end{figure}

\begin{sloppypar}
The cavity tuning mechanism consists of two sapphire strips with anisotropic relative permittivity $\varepsilon_\bot \approx 9$ and $\varepsilon_\parallel \approx 11$, relative to the crystallographic axis defined as the reference. Their dimensions are $12 \times 360 \times 2.56$\si{\milli\metre\cubed} and they are symmetrically placed parallel to the longitudinal sides moving simultaneously towards the centre (see Fig.~\ref{fig:cavity_assembly} bottom). These two parallel sapphire strips are displaced by a piezoelectric motor through a locomotive mechanism delivering a tuning resolution of better than \SI{100}{\Hz} in stable conditions (see Fig.~\ref{fig:cavity_assembly} bottom). The cavity frequency is changed by moving the sapphire strips towards or away from the split plane, while maintaining the parallelism to each other. The maximum tuning range for each cavity is about \SI{400}{\mega\Hz} corresponding to an axion mass range from $\sim \SI{21}{\micro\eV}$ to $\sim \SI{23}{\micro\eV}$. The cavity tuning system including the electromagnetic impact from the locomotive mechanism situated inside the cavity was designed to have no mode crossings for the axion mode over the entire tuning range. The maximum achieved scanning speed of CAST-CAPP is \SI{10}{\mega\Hz} / \si{minute} for each cavity, and therefore, its full tuning range of \SI{660}{\mega\Hz} can be covered within $\sim$\SI{1}{\hour}. The maximum speed is limited by the torque of the piezoelectric motors in cryogenic conditions. Following the CAST-CAPP protocol a given frequency is revisited multiple times for shorter periods rather than dwelling on it once for a long time interval. Thus the axion mass range is scanned horizontally with the sensitivity improving progressively. This gives us sensitivity not only to ALPs but also to axion mini-clusters, caustics and streams. In this work, the fast tuning mechanism allows primarily to quickly re-tune the cavities to a certain frequency in order to cross-check the nature of an outlier. 
\end{sloppypar}

The high mechanical $Q$-factor of the sapphire strips which are acting accidentally as a mechanical tuning fork are leading to a frequency modulation of the electromagnetic mode of interest. These ambient mechanical vibrations were reduced, increasing the measured effective $Q_{\text{L}}$ by a few \%, by introducing a vibration damping mechanism consisting of two quartz glass tubes filled with Teflon foils glued on the tuning sapphire strips. 

Each cavity assembly includes the cavity, the sapphire strips tuner, two RF couplers, one piezo-actuator, a locking device which fixes it to the CAST magnet bore, and one cryogenic amplifier placed at a few \si{\cm} distance from the edge of its cavity and with optimum orientation to the static B-field (see Fig.~\ref{fig:cavity_assembly} top). In addition, three out of the four cavities are equipped with Cernox temperature sensors. More specifically, one temperature sensor is installed in cavity 1, two in cavity 2 and one more in cavity 4. Each cavity is equipped with a weak coupler called ``injection port" and a near-critical coupler which we call ``main port". The near-critical coupling at cryo, for all four cavities was adopted in order to have optimum power transfer. The main port which is connected to a Low Noise Amplifier (LNA), is significantly under-critically coupled at room temperature in order to achieve near critical coupling at cryogenic temperatures with magnetic field within the available tuning range. This procedure was adopted because it was not possible to measure the coupling of the cavity directly and in situ under cryogenic conditions due to the permanently connected pre-amplifier. The strong coupling criterion is based on the experimentally-done observation that the coupling factor $\beta$  increases by a factor of $\sim 4$ for a tuneable cavity going from room to cryogenic temperatures. The coupling was characterized at warm by doing a reflection measurement at various frequencies around the resonance for both the very weakly coupled injection port and the strongly coupled main port. The length of the main coupler was adjusted accordingly in order to have four times less coupling than the critical one which is required in data-taking conditions neglecting the coupling at the injection port which was kept below $3\%$ both at cold and warm. The overall estimated uncertainty on the coupling from several performed measurements is below $10\%$.

For cryogenic temperatures, the $Q_{\text{L}}$ measured in transmission is about 20000 (at \SI{5.3}{\GHz}) which for near-critical coupling returns an unloaded Q of about 40000. However, this value of the unloaded Q could not be measured directly in our case since the near-critical coupled port was permanently connected to the low-noise pre-amplifier (no RF relay operating in a strong magnetic field is available). Thus, we are relying on the extrapolation from the $Q_{\text{L}}$ from warm to cold taking into account the Q-dependent change in the coupling factor $\beta$ and assuming negligible coupling on the other side (injection port) of the cavity. In addition, as a cross-check, we have the shape and height of the noise bump at cryo.

\begin{figure}[!htb]
\centering
\includegraphics[width=0.84\linewidth]{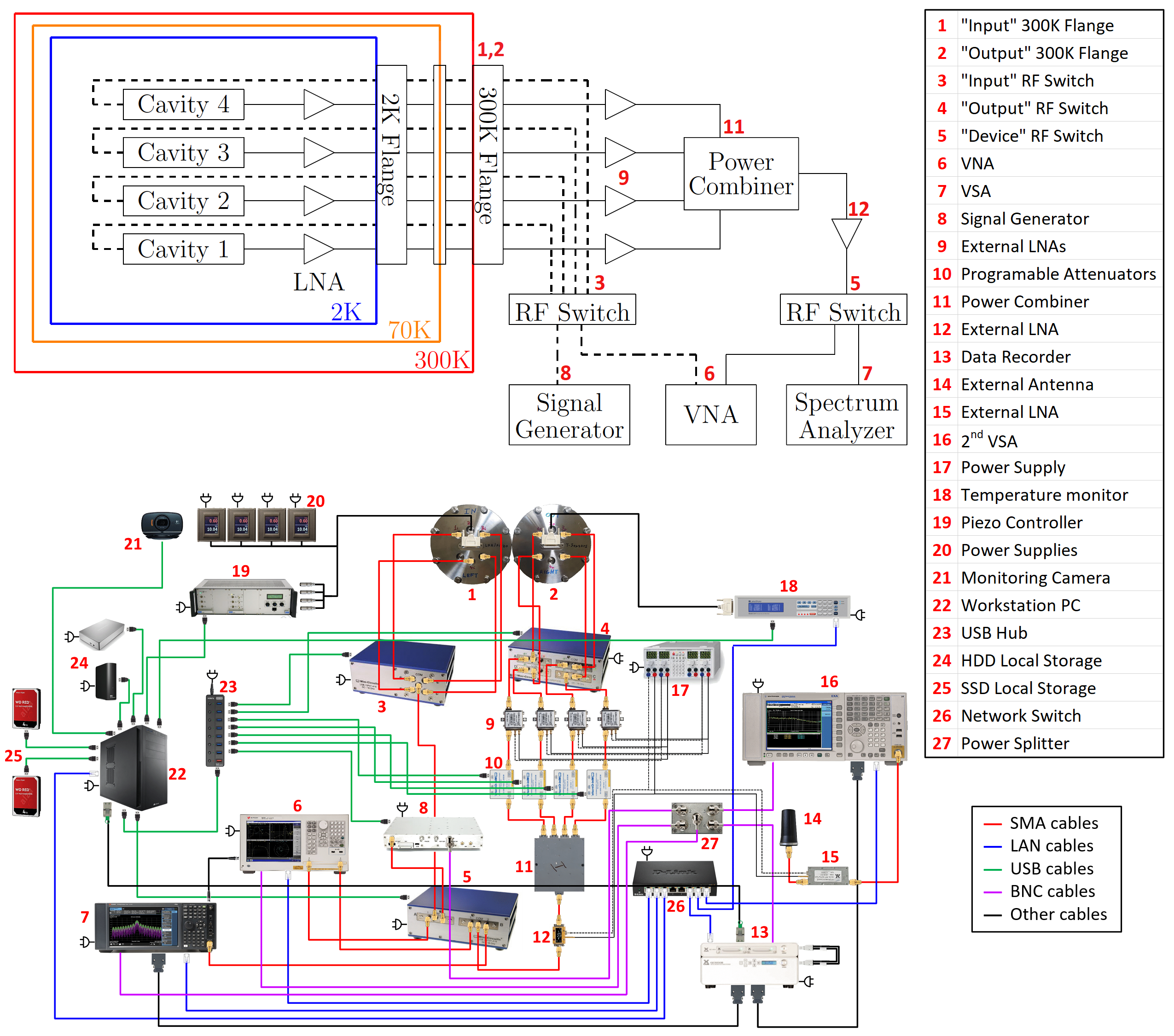}
\caption{Top: Simplified outline of the CAST-CAPP setup. The dashed and solid lines  correspond to connections in each cavity input and output port respectively. The cavities are actually aligned along the axis of their length as one after the other. Bottom: Detailed schematic of room temperature hardware connections. The numbers for the picture on the bottom refer also to the top.}
\label{fig:harware_schematic}
\end{figure}

The signal of each cavity at the cryogenic stage is amplified using an LNA with a gain of $\SI{39}{\dB}$ and then it is transferred from the cryogenic environment to room temperature through RF cables. Of course, these cables have certain frequency-dependent losses which have to be taken into account for the overall gain. An external signal amplification of about $\SI{22}{\dB}$ is made for each one of the four cavities with four room temperature amplifiers. The signals from the individual cavities are combined with a power combiner and transitioned to an RF switch sending the coherent signal to the measuring instruments. The simplified schematic of the setup both inside and outside the CAST magnet, is shown in Fig.~\ref{fig:harware_schematic}.

Using $N$ cavities, the SNR can be enhanced through hardware-implemented phase-matching, where, simultaneously acquired signals from several frequency-tracking cavities are combined coherently. In our case, the first stage amplification takes place before the signal combination, thus the SNR is improved linearly with the number of cavities \cite{jeong-2018}. This makes phase-matching the optimum configuration when using multiple cavities resulting in a much better performance compared to signal summation with undefined phases. A challenging part of the phase-matching of multiple cavities is the frequency alignment of all the cavities to the same resonant frequency. This has been achieved by adjusting the frequency of the cavities via tuning with the sapphire strips within \SI{10}{\kHz}, which is well below the allowed frequency matching tolerance of $\sim \SI{100}{\kHz}$ \cite{jeong-2018}. The amplitude of the resonance of the individual cavities was also adjusted within \SI{0.25}{\dB} using programmable phase-independent variable attenuators. The adjustment procedure uses the weakest signal as a reference and then the rest of the signals are adjusted via the phase-independent programmable step attenuators for each cavity path. The procedure of the frequency and amplitude alignment between the cavities has been performed for each data-taking step of \SI{200}{\kHz}. Finally, the output signals of each cavity have been adjusted using low-loss delay lines aiming for phase matching of the axion signal. The delay line is a frequency dependent phase shifter with a linear relation of phase-shift vs. frequency. This is what we are aiming at in order to achieve tracking of the phase-matching conditions over our frequency range without re-tuning the phase-shifting elements at each frequency step. More specifically, the phases were adjusted for coarse tuning by using short cable sections below \SI{5}{\cm} length and for the remaining difference continuously adjustable coaxial line stretchers. The hardware-based signal combination procedure is done at room temperature outside the magnet. By injecting to each cavity a pure sine-wave signal from a synthesizer and doing a complex addition of signals while taking into account the signal combiner properties, the linear increase on the SNR is observed in agreement with theory \cite{jeong-2018}.

The cavities' average physical temperature is about \SI{8}{\kelvin} or less, while the ambient temperature of the CAST magnet is $\SI{1.7}{\kelvin}$. The cavities are cooled along with the CAST magnet bores using liquid He ($\sim \SI{1.7}{\kelvin}$). This difference is due to contributions from the local heat dissipation by the LNAs being \SI{1.5}{\milli\watt} \cite{LNF} aggravated by the non-perfect thermal contact with the magnet bore. The temperature difference between LNA ON/OFF is $\sim \SI{2}{\kelvin}$. Piezo movement constitutes an additional heat source, however, impacting the overall cavity temperature by less than $\SI{0.4}{\kelvin}$.

The noise temperature was measured through the \SI{3}{\dB} signal generator method where a signal generator is used to inject a defined signal into the cavities aiming for a \SI{3}{\dB} difference between signal and no signal. It is noted that the LNA noise temperature is about \SI{2}{\kelvin} for our frequency range in cryogenic conditions and with proper orientation of the LNA with respect to the magnetic field. With this method, and considering a number of uncertainties for the losses from the cavities, we estimate an upper limit of the system noise temperature of \SI{10}{\kelvin}. Following our evaluation the system noise temperature is  $(9 \pm 1$) K.
Additionally, the loss of the connecting cable between the input of the amplifier and the output of the cavity at cryo was evaluated and is below \SI{0.1}{\dB}. Since this cable is also at cryo, its contribution to the system noise temperature is estimated to be safely below \SI{0.2}{\kelvin}.

Several ``intended" emitters in the experimental area, such as the \SI{5}{\GHz} Wireless Local Area Network (WLAN), were observed in our frequency range of operation. The path of the infiltration was tracked down as going via the RF amplifiers outside the cryostat. Two measures were taken: turning off part of the WLAN system in the \SI{5}{\GHz} band, and setting up a witness ``veto" channel to measure those time dependent signals online. The witness channel is made out of an additional Vector Signal Analyzer (SA) which is connected to an external quasi-omnidirectional antenna placed next to the CAST magnet. It is operating simultaneously with the SA connected to the cavities and at the same frequency band. Therefore, if a signal appears both in the cavities and in the witness channel at the same frequency it is rejected since an axion-related signal would only be generated inside the cavities which are permeated with magnetic field.

\begin{figure}[!htb]
\centering
\includegraphics[width=0.6\linewidth]{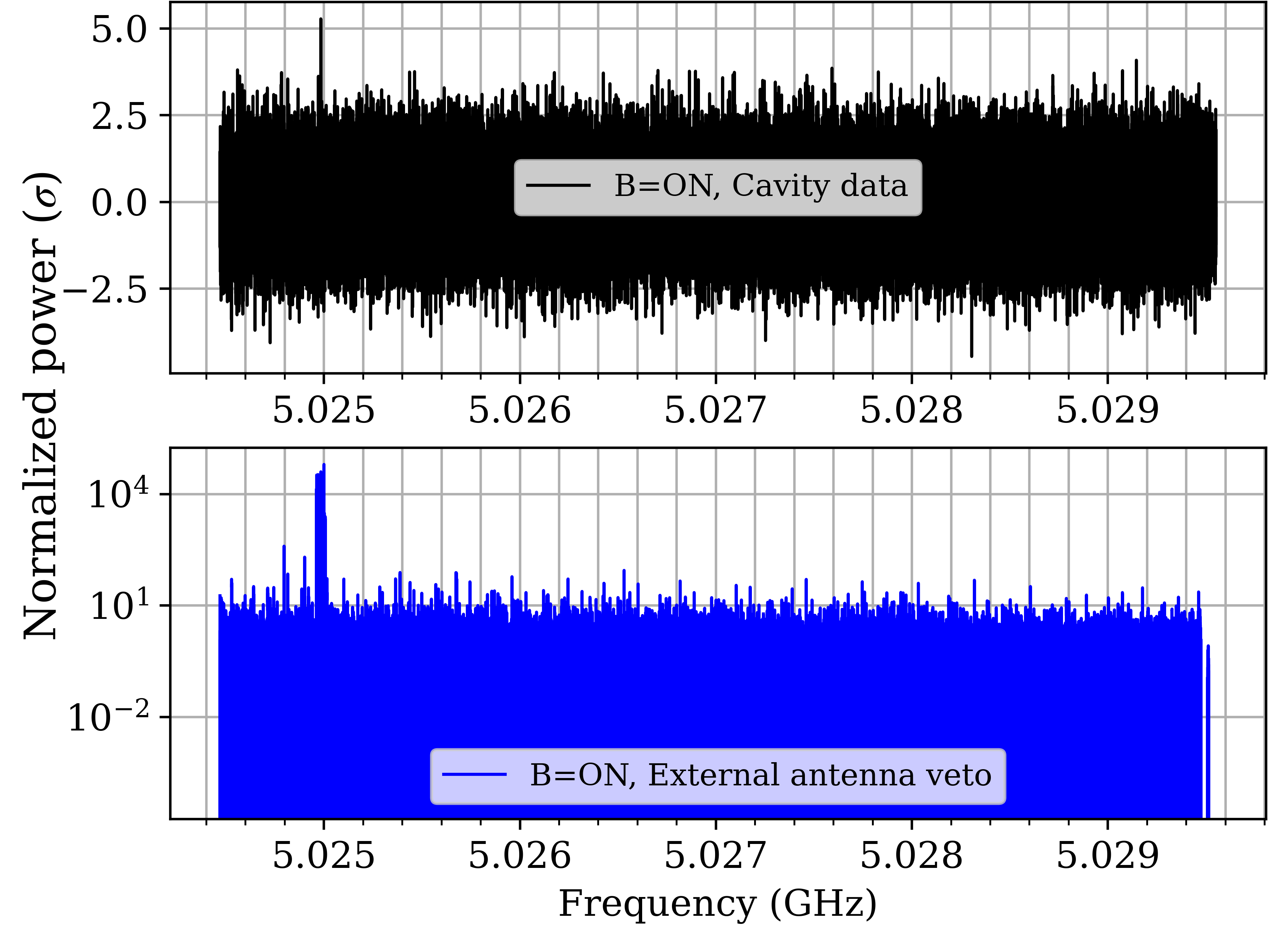}
\caption{An example of comparison of daily combined spectra between cavity (upper figure) and EMI/EMC channel (lower figure) both being equally processed. If a parasitic signal appears in both channels it is excluded from further consideration as an axion candidate.}
\label{fig:emi_comparison}
\end{figure}

The magnet vacuum vessel itself acts as a nearly-perfect Faraday cage. However, several vacuum coaxial feedthroughs reduce this ``perfect" shielding. In addition, an amount of electromagnetic interferences can make their way in the data acquisition (DAQ) chain though external amplifiers and electronics. However, there is a safety margin of \SI{30}{\dB}, i.e., as along as the Electromagnetic Interference (EMI)/Electromagnetic Compatibility (EMC) signals in the witness channel are less than \SI{30}{\dB} above the noise floor outside the magnet vessel, they can be excluded with the margin mentioned above. This limit comes from a measurement. We have intentionally radiated a well-defined signal in the CAST area via an additional antenna using the VNA as a signal generator. By analyzing the background through regular analysis we have verified that only strong signals beyond this amplitude may make their way into the cavity DAQ system. Therefore, the comparison between the two measuring channels allows us to directly control the impact of EMI/EMC parasites (see Fig.~\ref{fig:emi_comparison}).

\subsection{Data taking and analysis}
\label{S:2_2}

The DAQ chain is shown in Fig.~\ref{fig:harware_schematic}. The first part of the DAQ procedure is the measurement of all characteristic parameters required for analysis such as temperature, amplitude of resonant peak, Q-factor and center frequency for the cavity(ies). The data-recording from the SA begins with a streaming transfer of \SI{60}{\second} data blocks formatted as a real and imaginary part time domain trace. After each block the aforementioned parameters are re-measured in order to be used for quality checks as well as for averaging.

A calibration peak from a signal generator is regularly injected in order to verify that the whole DAQ chain and the cavities themselves are responsive to an actual axion signal, as well as to verify the analysis procedure.

During data-taking, 4 TB of data are produced per day, processed and uploaded to CERN's tape archiving system.

Both single and phase-matched cavity configurations are used in order to cover the maximum parameter space. More specifically, $62\%$ of the data were recorded with single cavities while $38\%$ with phase-matched cavities. The ideal tuning step size is of the order of the cavity frequency mode peak width $\sim \nu_0 / Q_{\text{L}}$. For our setup with ${\nu_0} \sim \SI{5}{\GHz}$ and $Q_{\text{L}} \sim 20000$, this results in a width of $\sim \SI{250}{\kHz}$. In order to account for possible variations in ${\nu_0}$ and $Q_{\text{L}}$ due to mechanical vibrations, the tuning step was set to \SI{200}{\kHz} whereas the maximum scanning speed of CAST-CAPP is about $\SI{10}{\mega\Hz}/\si{\minute}$. An integration time of \SI{60}{\second} was chosen for each of the \SI{200}{\kHz} data taking steps in every run.

From 12 September 2019 to 21 June 2021, it was scanned a total frequency range of \SI{660}{\MHz}, from \SI{4.774}{\GHz} to \SI{5.434}{\GHz}, in steps of \SI{200}{\kHz} for a total acquisition time of \SI{4123.8}{\hour}. The total frequency range corresponds to axion masses between \SI{19.74}{\micro\eV} and \SI{22.47}{\micro\eV}. The duty cycle of the data-taking campaigns was approximately \SI{20}{\hour} per day. Background measurements were also performed for a total of \SI{394.6}{\hour} without magnetic field.

From 18 November 2020 and on, a parallel channel has been used to measure EMI/EMC ubiquitous parasites yielding \SI{2140}{\hour} of data between \SI{4.798}{\GHz} and \SI{5.402}{\GHz}.

The raw data consist of in-phase (I) and quadrature (Q) components of time-domain (TD) voltage samples. In order to detect a potential axion signal the TD data are converted by a Fast Fourier Transformation (FFT) \cite{Cooley-1965} into frequency-domain (FD), where a peak is expected as the manifestation of a potential axion signal. The TD data are split into chunks and then are converted to FD by applying FFT to each individual chunk. Thus, we obtain I and Q samples in FD which are related to the RMS power via $P_{\text{rms}} = \left( {{I^2} + {Q^2}} \right)/\left( {2{Z_0}} \right)$, where $Z_0=\SI{50}{\ohm}$ is the SA input impedance. The number of samples in each FFT chunk is determined by the desired resolution bandwidth (RBW) $\delta \nu$ of the resulting processed spectra. The storage of the data allows a re-processing with FFT choosing a different RBW according to the needs of the analysis. We recall here that streaming axions should have a much smaller linewidth than the standard halo axions (of 5-7kHz) due to their very small velocity dispersion. At the same time, this format allows searching for short spikes and transient signals directly in the time domain trace in contrast with the frequency-domain which washes out any small and short-lasting signal.

To adjust $\delta \nu = \SI{50}{\Hz}$, we set the time length of each chunk as $\tau = 1 / \delta \nu = 0.02\,\text{s}$, which corresponds to $N = 0.02\,\text{s} \times 5 \text{ Mega samples/s} = 10^5$ samples per FFT chunk, given the sampling rate of 5 complex Mega samples/s. In order to obtain the processed spectra out of each $t=\SI{60}{\second}$ measurement $t/\tau=3000$ FFT chunks are averaged. The acquisition bandwidth of each processed spectrum is \SI{5}{\MHz}, being deliberately much wider compared to the $\approx \SI{250}{\kHz}$ cavity resonance width.

A number of selection criteria, as shown in Table \ref{tab:discarded-datafiles}, are applied on each processed block before the analysis procedure takes place. These criteria are applied to ensure exclusion of undesired non-systematic effects from mechanical vibrations or possible EMI/EMC interferences, which could otherwise influence the center frequency of the resonant peak, its amplitude and the measured $Q_{\text{L}}$. These characteristic parameters are measured by the VNA before and after each single \SI{60}{\second} measurement block. 

For the data taken with $B=\SI{8.8}{\tesla}$, the total number of processed blocks is 262491 out of which $\sim 4.4\%$ did not pass the quality checks. For the data without magnetic field, 461 blocks out of 24138 were discarded corresponding to a rejection factor of $\sim 1.9\%$.

\begin{table}[htbp]
\centering
\caption{Data qualification criteria. ``Before" and ``after" labels indicate the measurements before and after each measurement block, respectively. Criteria 1-4 apply to all measurements while 5-8 refer only to phase-matched data taking.\vspace{1em}}
\begin{tabular}{l l l}
\hline
\textbf{Nr.}&\textbf{Parameters} & \textbf{Criteria}  \\
\hline
1   &   Resonance frequency stability      & $\Delta \nu_0 <$ \SI{100}{\kHz} \\ 
2   &   Resonance amplitude variation      & $\Delta A_0 <$ \SI{3}{\dB}    \\ 
3   &   Loaded quality factor       & $10^3 < Q_{\text{L}} < 4 \times 10^4$       \\ 
4   &   Loaded quality factor shift & $\Delta Q_{\text{L}} < 7 \times 10^3$       \\
5   &   Frequency mismatch   & $<$ \SI{20}{\kHz} (before)       \\
6   &   Frequency mismatch   & $<$ \SI{80}{\kHz} (after)       \\
7   &   Amplitude mismatch   & $<$ \SI{1}{\dB}                 \\
8   &   Temperature mismatch & $<$ \SI{3}{\kelvin}             \\ 
\hline
\end{tabular}
\label{tab:discarded-datafiles}
\end{table}

The analysis procedure is mainly based on widely accepted methods \cite{ADMX-2001, HAYSTAC-2017, ADMX-2021}. Nevertheless, we make a few modifications to adapt this procedure to our experimental conditions. We start with the removal of the noise baseline of the processed spectra using a Savitzky-Golay (SG) smoothing filter \cite{Savitzky-1964} with window length $W = 1000$ bins and a polynomial of order $d = 4$. The processed spectra are divided by their SG filter output and subtracted 1 to obtain a mean $\mu = 0$ with standard deviation $\sigma = (\delta \nu \cdot t)^{-1/2} \approx 0.018$ of a Gaussian distribution. This sample distribution is a result of averaging 3000 FFT chunks as dictated by the central limit theorem, whereas the samples of each chunk, in power units, obey a $\chi^2$ distribution. We remove the intermediate frequency (IF) filter roll-off from the two edges of these flattened spectra which correspond to 20\% of the samples in total. Therefore, each flattened spectrum has $10^5$ bins.

The flattened spectra are scanned for IF interferences which occur as unexpectedly high-amplitude bins on the same index of most of the flattened spectra regardless of the cavity resonance frequency. First, all flattened spectra are divided into three groups, and each group is averaged according to the IF bin index. Second, we flag the bins that exceed 5$\sigma$ with $\sigma = (M \, \delta \nu_P \, t)^{-1/2}$, where M is the number of averaged spectra in each group. Finally, we compare the flagged IF bin index list. If an IF bin is flagged in at least 2 groups, we discard that particular IF bin plus the 2 adjacent bins of all flattened spectra from the subsequent steps of the analysis procedure, this is the case for 6 out of the $10^5$ bins.

Each flattened spectrum is scaled by $P_{\text{noise}} / P_{\text{axion}}$ in order to attain an axion SNR as given in Eq.~\eqref{eq:snr}, where $P_{\text{axion}}$ is the axion signal power curve across the bandwidth of each spectrum as defined by Eq.~\eqref{eq:axion-power}.

The combined spectrum is generated by aligning the scaled spectra on the RF axis and taking the weighted average of the scaled spectra bin amplitudes that correspond to the same RF bin of the combined spectrum with weights equal to inverse bin variances, as determined by maximum likelihood (ML) estimation. The amplitude $p_k$ of each combined spectrum bin of index $k$ is normalized by dividing it by its standard deviation $\sigma_k$, resulting to normalized samples with a normal distribution of mean $\mu = 0$ and standard deviation $\sigma = 1$, except for the prospective bin that contains axion power which must be a sample of a normal distribution with mean $(\sigma_k)^{-1}$ and $\sigma = 1$. The resolution bandwidth of the combined spectrum remains $\delta \nu_C = 50$ Hz. 

In order to increase the SNR of an axion signal of linewidth $\Delta \nu \approx \SI{7}{kHz}$ in the lab frame, the adjacent 28 bins in the combined spectrum are averaged with their ML weights, obtaining a rebinned spectrum with increased bandwidth resolution $\delta \nu_R = \SI{1.4}{kHz}$. 

\begin{figure}[!htb]
\centering
\includegraphics[width=0.8\linewidth]{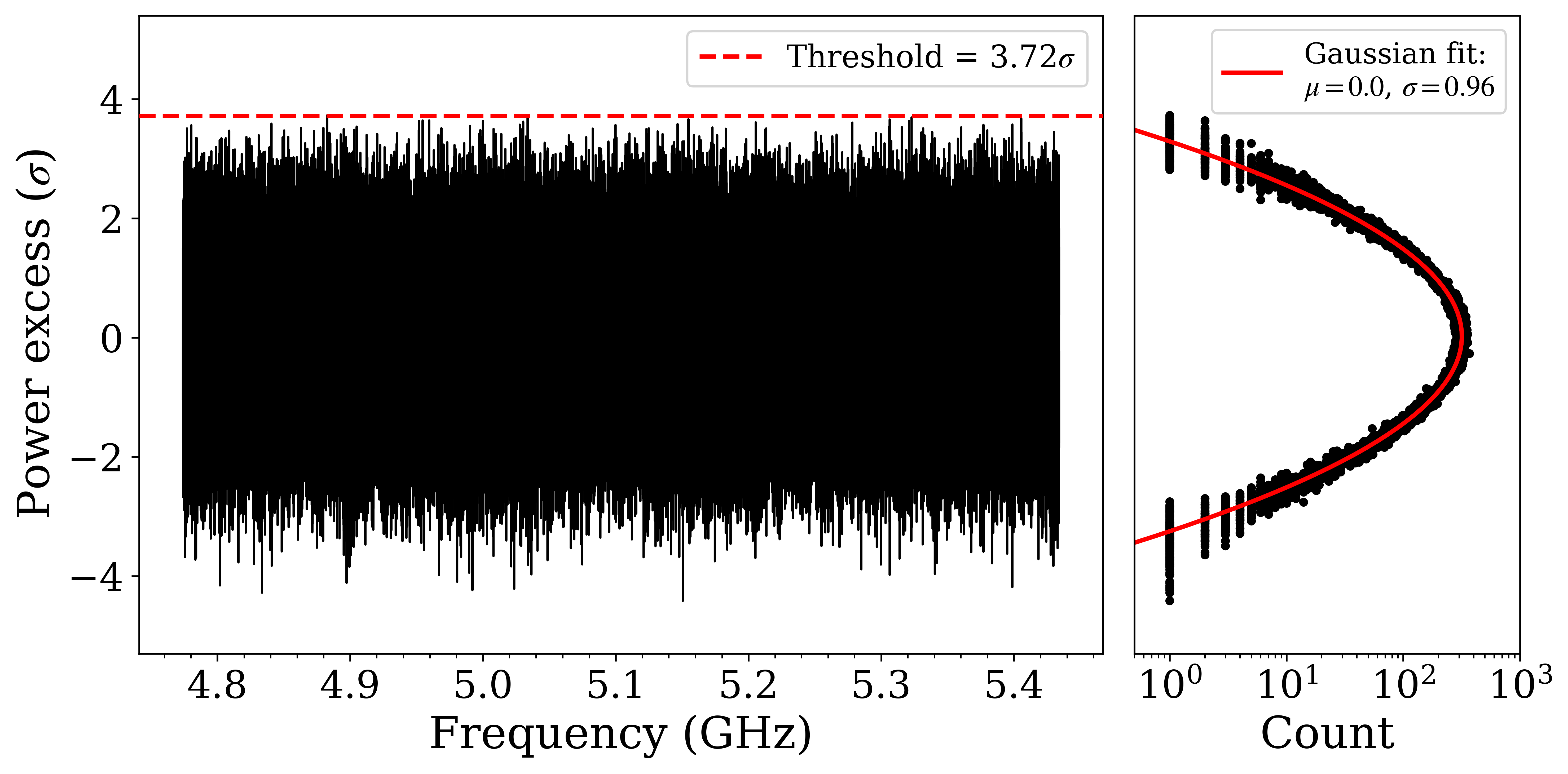}
\caption{Grand spectrum (left) and its projected noise distribution (right). All the grand spectrum bins are below the selection threshold (red dashed line) after the rescanning procedure which shows that all the axion candidates are excluded.}
\label{fig:grand_spectrum_snr_together}
\end{figure}

Finally, the ``grand spectrum" is computed by convolving the rebinned spectrum with the kernel equal to the axion lineshape in lab frame defined by Eq.~\eqref{eq:axion-pdf}. This leads to an unchanged bandwidth resolution and number of samples of the grand spectrum. The normalized grand spectrum is generated by dividing each bin's power by its standard deviation $p_k^G / \sigma_k^G$, where the noise distribution should have $\mu = 0$ and $\sigma = 1$. However, the standard deviation is reduced to $\sigma = 0.74$, due to the negative correlations among adjacent bins introduced by the SG filter. To quantify the covariance among consecutive grand spectrum bins, a Monte Carlo simulation of $10^4$ processed spectra has been constructed using sample noise baselines from real data. Elements of the covariance matrix resulting from this simulation are used to compute the ML estimate of the standard deviation of the grand spectrum bins. The so derived normalized grand spectrum is described by $\sigma=0.96$. Fig.~\ref{fig:grand_spectrum_snr_together} shows the grand spectrum and its noise distribution.

\subsection{Analysis results}
\label{S:2_3}

We search the grand spectrum for bins that could show an axion DM signal indicated by a positive power excess. Applying the SG filter on the analysis procedure reduces the grand spectrum SNR by $\eta=0.717$ as derived by Monte Carlo simulations of $10^4$ experiments with axion signal injections. To search for the KSVZ axion, accepting the target SNR$_{\text{T}} = 5 \sigma$ and confidence level of 90\%, the candidate selection threshold is set accordingly to $3.72 \sigma$. The grand spectrum bins which exceed this threshold are then flagged as rescan candidates. We have in total 60 bins out of $\approx$ 472k that are flagged for rescanning. 

Following a predefined procedure, every candidate spectrum is compared to the simultaneously and independently measured spectrum from the second channel with the external antenna which is sensitive to the same frequency and searches for ambient EMI/EMC parasites. A total of nine out of the $60$ outliers 
were discarded from further consideration with three of them being rejected following background measurements without a magnetic field. Furthermore, 11 outliers were blind calibration signals deliberately injected into the cavities. The next step is to rescan a narrow band around each of the 40 remaining outliers with the same and different cavities to increase statistics, but also for consistency checks. If a candidate is persistent enough, an extra rescanning is performed by tuning to a higher order resonant mode that does not couple to axions. Finally, if an outlier also passes this elimination step, a rescan at a different magnetic field strength takes place. 

No outlier has reached these two last predefined steps because all 40 outliers were discarded during rescanning. We can thus compute the excluded coupling strength $|g_{a\gamma\gamma}|$ for each bin of the grand spectrum $|g_{{a\gamma\gamma}}| = |g_{{a\gamma\gamma}}^{\text{KSVZ}}| \, \sqrt{\text{SNR}_{\text{T}} / \text{SNR}}$. 

\begin{figure}[!htb]
\centering
\includegraphics[width=1\linewidth]{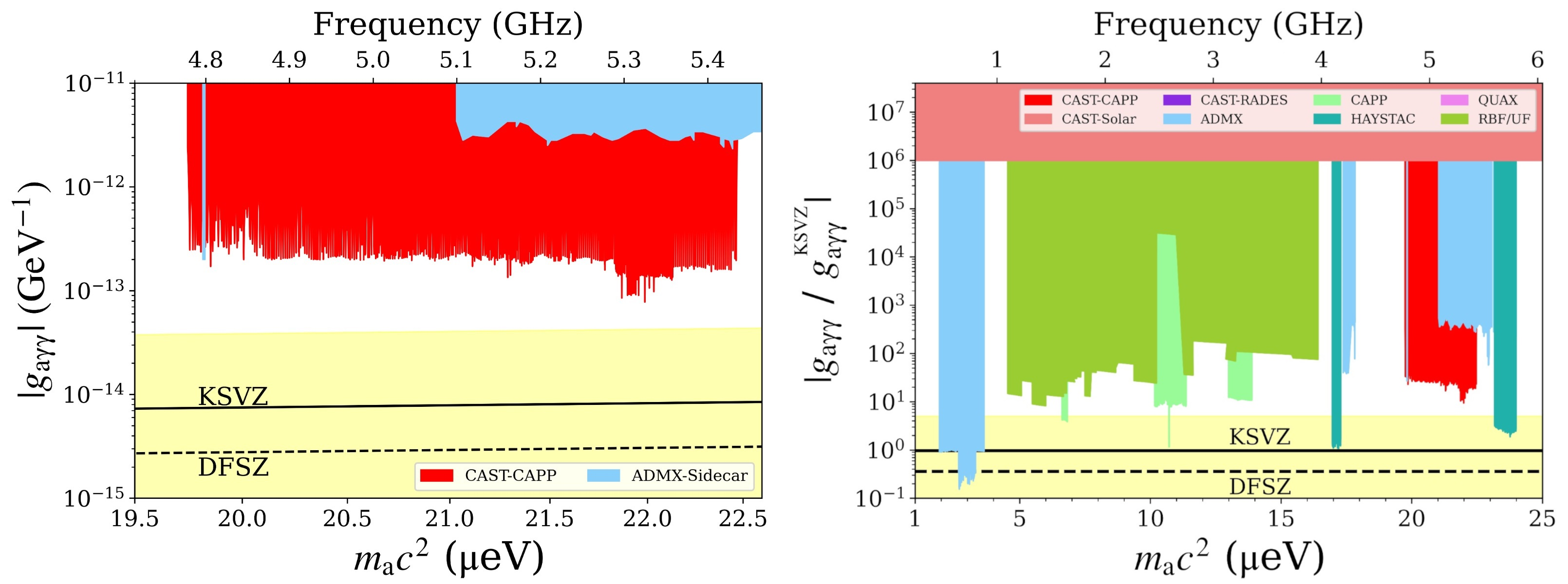}
\caption{CAST-CAPP exclusion limit on the axion-photon coupling as a function of axion mass at $90\%$ confidence level (left), and compared to other axion search results \cite{RBF-1987, UF-1990, CAST-2017, HAYSTAC-2017, ADMX-Sidecar-2018, CAPP-2020-1, CAPP-2020-2, ADMX-2021, ADMX-2021-1, HAYSTAC-2021, CAPP-2021} within the mass range 1-25~\si{\micro\eV} (right). The higher sensitivity around the \SI{22}{\micro\eV} region, is due to the predominant use of phase-matching around this mass range.}
\label{fig:exclusion_plot}
\end{figure}

The exclusion of axion-photon coupling at 90\% confidence is shown in Fig.~\ref{fig:exclusion_plot}, where the estimated overall uncertainty calculated from the individual uncertainties shown in Table \ref{tab:capp-parameters}, is $10\%$. We exclude previously unexplored new parameter space for axion masses between \SI{19.74}{\micro\eV} and \SI{22.47}{\micro\eV}.

\begin{table}[htbp]
\centering
\caption{CAST-CAPP parameters and related uncertainties used for the analysis. Note that the loaded quality factor is given for the frequency of \SI{5.3}{\GHz}.\vspace{1em}}
\begin{tabular}{l l l l}
\hline
\textbf{Parameters} & \textbf{Explanation} & \textbf{Values} & \textbf{Uncertainty}  \\
\hline
$B$     &   Static dipole magnetic field    & \SI{8.8}{\tesla}              & $10^{-3}$  \\ 
$V$     &   Cavity volume                   & \SI{224}{\centi\meter\cubed}  & \SI{0.1}{\centi\meter\cubed}  \\ 
$C$     &   Form factor                     & 0.53                          & $10\%$  \\ 
$\beta$ &   Main port coupling factor       & 1                             & $0.3$  \\
$Q_{L}$ &   Loaded quality factor           & 20000                         & $3\%$  \\
$T_S$   &   System noise temperature        & \SI{9}{\kelvin}               & \SI{1}{\kelvin}  \\
$\eta$  &   Signal attenuation coefficient  & 0.717                         & 0.01  \\
\hline
\end{tabular}
\label{tab:capp-parameters}
\end{table}

\section{Discussion}
\label{S:3}

We report here the first results of the CAST-CAPP detector searching for galactic DM axions. CAST-CAPP is a haloscope consisting of four individual resonant cavities inserted in one bore of the CAST dipole magnet. 

A previously unexplored parameter space has been scanned to extend the axion search towards larger rest mass values from \SIrange{19.74}{22.47}{\micro\eV}. This required higher frequencies and thus a smaller cavity size and a tuning mechanism with no mode-crossings for the mode of interest over the entire frequency range. This is a particular strength of the cavity design. In this mass range, we exclude axion-photon coupling for virialized galactic axions down to $g_{a\gamma\gamma} = 8 \cdot 10^{-14} ~\text{GeV}^{-1}$ at 90\% confidence level. 

In addition, the following novelties, of potential importance also for future DM axion searches, have been instrumented:
\begin{itemize}
    \item Four identical cavities have been coherently combined through the phase-matching technique increasing significantly the SNR. Combining signals from individual cavities is shown to be feasible and still unique in DM axion search. This can be extended in future large-scale axion haloscopes with a large number of small cavities.
    \item The successful scan of a  significant mass range showed that this experiment is at the cutting-edge of cavity tunability in axion research. This was achieved thanks to the unique design of the locomotive tuning mechanism and the cavity geometry. With an upgrade of the piezoelectric motors at cryo, the tuning range can even further be extended to  \SI{1}{\GHz} and slightly beyond (\SI{4.6}{\GHz} - \SI{5.8}{\GHz}), shown by simulation \cite{Miceli-haloscope} and also demonstrated on the bench at room temperature.
    \item The fast-scanning technique includes a fast change of resonance frequencies (10 MHz / min) between \SI{4.774}{\GHz} and \SI{5.434}{\GHz}, which, combined with a  high sensitivity, allows for \SI{60}{\second} short acquisition intervals for each \SI{200}{\kHz} tuning step. This also permits to quickly re-tune the cavities to a frequency of interest, for instance to investigate a stable and reproducible outlier.
    \item The raw data which are recorded in each interval consist of a ``real and imaginary part" time domain trace allowing for easy analysis of the traces for transient event search. Together with the fast-scanning technique, it permits a simultaneous search for halo DM axions and axion-caused electromagnetic transients originating for example from mini-clusters or streams \cite{vogelsberger-2011,Tkachev-1991, Kolb-1993}. It is challenging to separate non-stationary signals from axion streams against transient-type EMI/EMC perturbations. The very first adequate approach in this case is the comparison with the reference antenna output in the hall.
\end{itemize}

\noindent Note added in Proof: \\ During the review process another publication \cite{CAPP-4} appeared overlapping partly with our low axion mass range.  

\section*{Data availability}

The source data and code underlying the plots in this paper are available from the corresponding authors upon request.

\bibliographystyle{elsarticle-num-names}
\bibliography{bibliography.bib}

\section*{Acknowledgements}

\begin{sloppypar}
We wish to thank our colleagues at CERN, in particular Thomas Schneider and Miranda Van Stenis from the glass lab, as well as Manfred Wendt and the whole team of the CERN Central Cryogenic Laboratory for their help and overall support. We  highly appreciate the help of the dental expert Dr. Sebastian Engelhardt, Satigny in a post data taking repair / upgrade. We would also like to thank Spyridon Maroudas for creating the first two figures of this paper. This work has been supported by the Greek General Secretariat for Research Innovation (GSRI), the Institute for Basic Science (IBS) under the Project Code No. IBS-R017-D1 of the Republic of Korea, the Spanish Agencia Estatal de Investigacion (AEI) and Fondo Europeo de Desarrollo Regional (FEDER) under project FPA-2016-76978-C3-2-P and PID2019-108122GB-C33, by the CERN Doctoral Studentship programme, the European Research Council (ERC), the DFG Research Training Group Programme 2044 “Mass and Symmetry after the Discovery of the Higgs Particle at LHC”, MSE (Croatia) and Natural Sciences and Engineering Research Council of Canada. We acknowledge support through the ERC under grant ERC2018-StG-802836 (AxScale project) and ERC-2017-AdG-788781 (IAXO+ project). Part of this research is co-financed by Greece and the European Union (European Social Fund - ESF) through the Operational Programme ``Human Resources Development, Education and Lifelong Learning" in the context of the project ``Strengthening Human Resources Research Potential via Doctorate Research – 2nd Cycle" (MIS-5000432), implemented by the State Scholarships Foundation (IKY). Part of this work was performed under the auspices of the US Department of Energy by Lawrence Livermore National Laboratory under Contract No. DE-AC52-07NA27344.
\end{sloppypar}

\section*{Author contributions}
G.C., F.C., S.A.C., W.C., H.C., J.C., H.F., W.F., M.K., C.K., J.L., S.L. M.M., L.M., K.O., Y.K.S., T.V., S.Y. and K.Z. conceived, designed and built the CAST-CAPP detector. F.C., A.G., J.M.L., M.M., L.M., K.O., Y.K.S. and T.V. installed, and operated the detector. F.C., H.F., W.F., M.M., K.O., T.V., K.Z. organized the data-taking runs and general operation of the experiment, coordinated, monitored, processed and analysed the detector data. F.C., S.A.C., H.F., W.F., M.K. M.M., K.O., G.C., Y.K.S., K.Z. wrote the manuscript. All authors contributed to the data-taking shifts or the operation and maintenance of the experiment at a whole.

\section*{Competing interests}

The authors declare no competing interests.

\end{document}